\documentclass[12pt,preprint]{aastex}

\shorttitle{A New Detached M Dwarf Eclipsing Binary}
\shortauthors{Creevey et al.}
\begin{document}
\title{A New Detached M Dwarf Eclipsing Binary}
\author{
O.L. Creevey\altaffilmark{1,2,3}, 
G.F. Benedict\altaffilmark{4}, 
T.M. Brown\altaffilmark{1}, 
R. Alonso\altaffilmark{3}, 
P. Cargile\altaffilmark{4}, 
G. Mandushev\altaffilmark{5}, 
D. Charbonneau\altaffilmark{6},  
B.E. McArthur\altaffilmark{4},  
W. Cochran\altaffilmark{4}, 
F.T. O'Donovan\altaffilmark{7}, 
S.J. Jim\'enez-Reyes\altaffilmark{3}, 
J.A. Belmonte\altaffilmark{3},
D. Kolinski\altaffilmark{1}}

\email{creevey@hao.ucar.edu}
\email{fritz@astro.as.utexas.edu}
\email{timbrown@hao.ucar.edu}
\email{ras@iac.es}
\email{p.cargile@mail.utexas.edu}
\email{gmand@lowell.edu}
\email{dcharbonneau@cfa.harvard.edu}
\email{mca@barney.as.utexas.edu}
\email{wdc@shiraz.as.utexas.edu}
\email{francis@caltech.edu}
\email{sjimenez@iac.es}
\email{jba@iac.es}
\email{kolinski@hao.ucar.edu}

\altaffiltext{1}{High Altitude Observatory, National Center for Atmospheric Research, Boulder 80307, CO USA}
\altaffiltext{2}{Universidad de La Laguna, 38206 La Laguna, Tenerife, Spain }
\altaffiltext{3}{Instituto de Astrof\'isica de Canarias, E-38200 La Laguna, Tenerife, Spain}
\altaffiltext{4}{McDonald Observatory, University of Texas, Austin, TX 78712 USA}
\altaffiltext{5}{Lowell Observatory, Flagstaff, AZ 86001 USA}
\altaffiltext{6}{Harvard-Smithsonian Center for Astrophysics, 60 Garden St., Cambridge, MA 02138 USA}
\altaffiltext{7}{California Institute of Technology, 1200 E. California Blvd., Pasadena, CA 91125 USA}

\begin{abstract}
We describe a newly-discovered detached M-dwarf eclipsing binary system. 
This system was first observed by the TrES network during a long term 
photometry campaign of 54 nights.  
Analysis of the folded light curve indicates two very similar 
components orbiting each other with
a period of 1.12079 $\pm$ 0.00001 days.  
Spectroscopic observations 
with the Hobby-Eberly Telescope show the system to consist of two
M3e dwarfs in a near-circular orbit.  
Double-line radial velocity amplitudes, 
combined with the orbital inclination derived from light-curve fitting, yield
M$_{total}$ = 0.983 $\pm$ 0.007 M$_{\odot}$, with component masses of 
M$_1$=0.493 $\pm$ 0.003 M$_{\odot}$ and M$_2$=0.489 $\pm$ 0.003 M$_{\odot}$. 
The light-curve fit yields component radii of R$_1$=0.453 $\pm$ 0.060 R$_\odot$ and
R$_2$=0.452 $\pm$ 0.050 R$_\odot$.
Though a precise parallax is lacking, broadband VJHK colors and spectral typing
suggest component absolute magnitudes of M$_V$(1) = 11.18 $\pm$ 0.30 and 
M$_V$(2) = 11.28 $\pm$ 0.30. 

\end{abstract}


\section{Introduction}
\begin{figure*}
\plotone{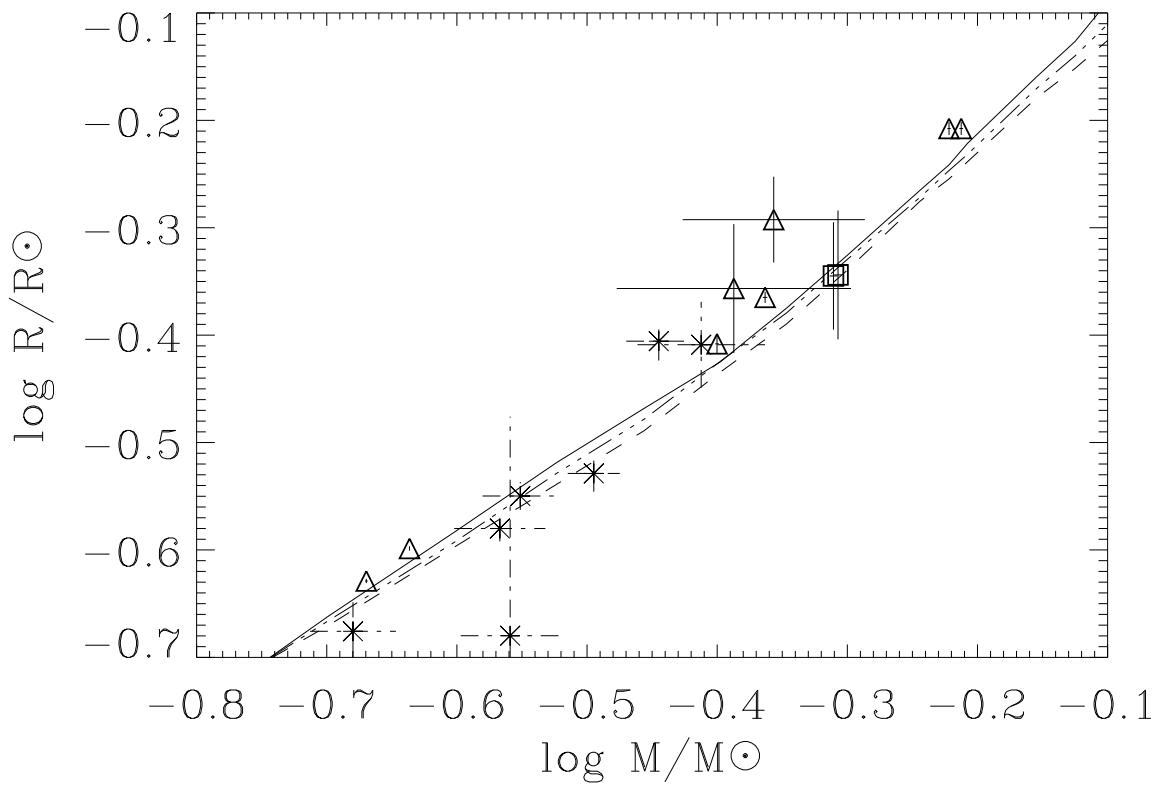}
\caption{Mass-Radius relation showing the M-dwarf 
component parameters (triangles).  
Our binary system is represented by squares.  
We also represent some other M dwarf stars, whose companions are 
F or G MS stars (Bouchy et al. 2005; Pont et al. 2005)
We show theoretical models from Baraffe (1998), indicating an age of 10 
(continuous), 5 (dash-dotted) and 1 Gyr (dashed), 
corresponding to [Z]=0,  Y=0.275.  
 }
\label{gr_mr}
\end{figure*}

Although low mass binary stars are the most abundant stars in the galaxy 
(Henry et al. 1999), 
their intrinsic faintness inhibits their detection and study. 
Non-contact eclipsing binary M dwarf systems have 
great value, as these systems allow accurate estimates
of the most basic stellar parameters: mass and radius.  
Only four\footnote{We refer specifically to binaries where both components are M Dwarfs.  There have been, however, a number of M stars whose companion is an F or G MS star (e.g. Pont et al., 2004), see Figure \ref{gr_mr}} such systems are known and have been studied in detail;  
YY Gem (Bopp 1974; Leung \& Schneider 1978), 
CM Dra (Lacy 1977; Metcalfe et al. 1996; Kozhevnikova et al. 2004), 
GJ 2069A (Delfosse et al. 1999; Ribas 2003),
and OGLE BW03 V038\footnote{This is a very close although still detached 
system.} (Maceroni \& Montalban, 2004).
The observed properties of each of these systems
present discrepancies with the theory of low-mass stellar objects;
neither the observed mass-radius nor mass-luminosity 
relations are well represented
by existing models (Benedict 2000); see Figure \ref{gr_mr}.
The problem most likely lies in the shortcomings of the physical models, 
owing to the lack of understanding of the complex atmospheres 
of such low-mass objects (Baraffe et al. 1998).  
Enlarging the small existing sample of such systems is therefore desirable, 
to allow more detailed comparisons between observations and the theory
of these ubiquitous, interesting, and complex objects.
Here we report preliminary analysis of a fourth such low-mass eclipsing binary.

\section{Observations}
\subsection{Photometric Observations}
The recently discovered spectroscopic binary,  TrES--Her0-07621
($\alpha$=16$^h$50$^m$20.7$^s$, $\delta$=+46$^{\circ}$39$'$01$''$ (J2000), $V$=15.51 $\pm$ 0.08)  
was first identified through an analysis of photometric time series from 
the TrES (Trans-Atlantic Exoplanet Survey) network.  
This network consists of three telescopes:
{\it STellar Astrophysics and Research on Exoplanets} (STARE,\footnote{Observatorio del Teide, Tenerife, Spain}
 Brown \& Charbonneau, 1999), 
{\it Planet Search Survey Telescope} 
(PSST\footnote{Lowell Observatory, AZ, USA}, Dunham et al. 2004), 
and {\it Sleuth}\footnote{Palomar Observatory, CA, USA. \tt http://www.astro.caltech.edu/$\sim$ftod/tres/sleuth.html }. 
The telescopes are similar in their characteristics, 
with apertures of 10cm, 2048$\times$2048 pixel CCD detectors 
and fields of view of 6$^{\circ}\times$6$^{\circ}$.

TrES collects long-term time-series photometry in one filter. 
The photometry run in question spanned 54 days, 
beginning May 6 2003, and was observed in a band roughly
equivalent to Harris R at a cadence 
of 1 image every 2 minutes.  
The images were reduced and calibrated by an automatic package 
developed specifically for these data.  
TrES--Her0-07621 was observed by both STARE and PSST, 
but the latter time series proved significantly noisier.
We therefore analyzed only the STARE lightcurve.  The R magnitude is 14.42 
with each point having a formal accuracy of 0.04 mag rms.  
This lightcurve 
contains 8781 data points, obtained in 
309.5 hours over 54 days, giving a duty cycle of 
23.8\%{\bf \footnote{The data are available via STARE website
{\tt http://www.hao.ucar.edu/public/research/stare/stare.html} } }.

A high-SNR peak in the time series' frequency spectrum at 1.79 cycles per day 
initiated the study of TrES--Her0-07621. 
Folding the star's light curve with a period of 1.1208 d showed it to be
an eclipsing binary.
The light curve also displays sinusoidal out-of-eclipse variations 
near the photometric period.
The star's infrared colors from the 
2MASS\footnote{Two Micron All Sky Survey: 
University of Massachusetts and the Infrared Processing and Analysis 
Center/California Institute of Technology  
\tt http://irsa.ipac.caltech.edu/cgi-bin/Gator/nph-dd } 
catalog are quite red, (Table \ref{facts}), and the 
USNO-B\footnote {SIMBAD, operated at CDS, 
Strasbourg, France; the NASA/IPAC Extragalactic Database (NED) 
and supported by JPL, California Institute of Technology, 
{\tt http://www.nofs.navy.mil/data/fchpix/}} catalog
 shows a significant proper motion. 
Taken together, these facts suggested that the object is a binary
M dwarf, with substantial levels of magnetic activity driven by the
rapid, tidally-locked rotation of the component stars;  this motivated 
further study.

\subsection{Spectroscopic Observations}
In September 2004 we obtained spectroscopic observations
of TrES--Her0-07621 using the High Resolution Spectrograph (HRS, Tull 1998) on the 
Hobby-Eberly Telescope (HET).
We secured
measurements at 4 epochs;  
each epoch contained three separate exposures taken over 
approximately one hour --- giving a total of 12 spectra.  
The analysis was carried out with standard IRAF (Tody 1993) {\tt echelle} 
and {\tt rv} package tools, including {\tt fxcorr}.  
We cross-correlate TrES-Her0-07621 with
an M2 dwarf (Gl 623) template and extract velocities for both components
at four distinct phases.  We adopted a radial velocity for the Gl 623 primary
of -29.2 km s$^{-1}$, given the orbital phase at which the template was 
secured and a  systematic velocity, V$_{sys}$ = -27.5 km s$^{-1}$, 
from Marcy \& Moore (1989).  
The HRS utilizes two CCDs covering the blue and red spectral regions.
The data from each chip were analyzed independently, resulting in two velocity
estimates.  A third velocity estimate was obtained by cross-correlating an
artificial H-alpha emission template with the H-alpha emission line found
 in each exposure.  
Given the large orbital velocities, there was no 
blending of correlation peaks at any phase.
The three velocities (blue, red, H$\alpha$) are obviously not independent
determinations, but do provide an estimate of our internal error.

\section{Analysis}
\begin{figure*}
\plotone{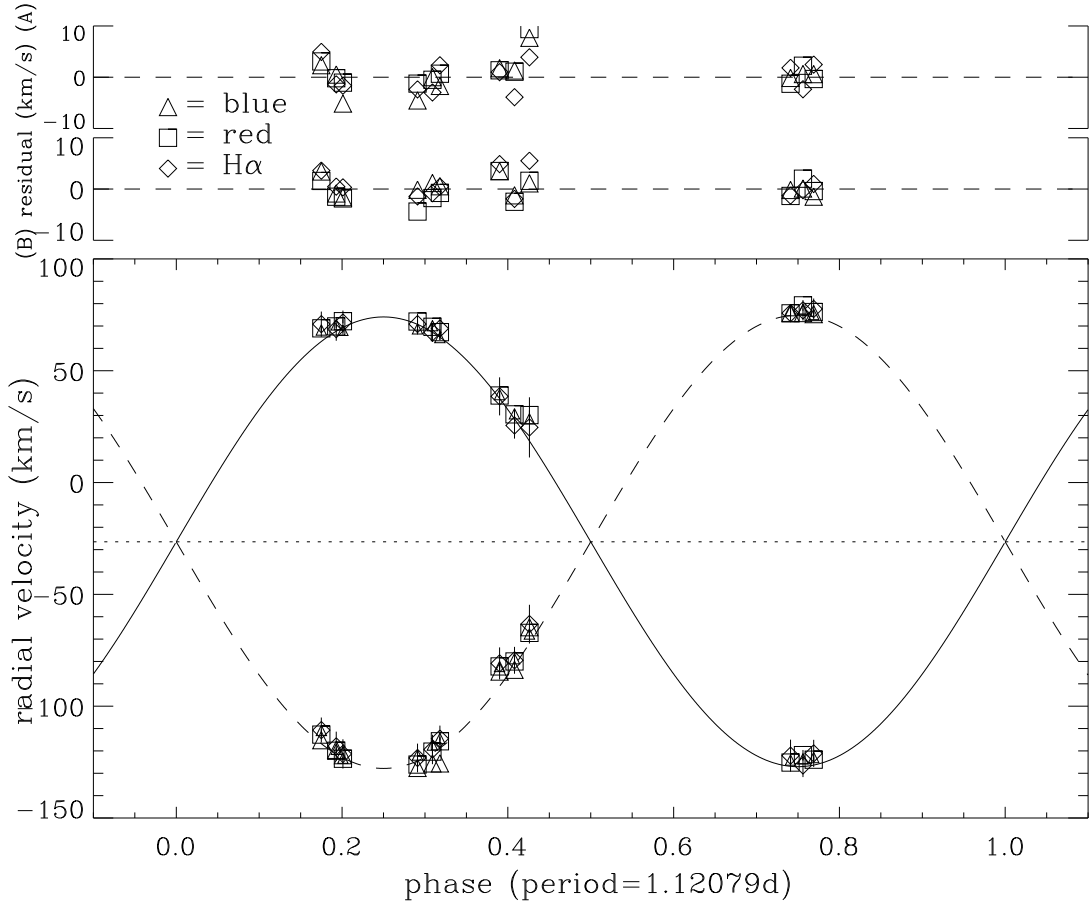}
\caption{Phased radial velocity curve of TrES-Her0-07621 with residuals; 
data observed by the HET.  
A total of four nights were obtained, the system was observed for 
one hour every night at 20 minutes per exposure.  The Doppler displacement
was measured in two different wavelengths regions and at H$\alpha$.}
\label{gr_het}
\end{figure*}

Figure \ref{gr_het} shows the component velocities plotted against 
photometric phase, while figure \ref{gr_fit_res} (top panel) shows the 
folded photometric light curve.
It is evident from the
nearly-symmetrical and sinusoidal radial velocity variation and 
from the highly symmetrical light curve
that the orbit is nearly circular, and that the component masses
and surface brightnesses are similar.

An initial period analysis of the entire STARE lightcurve 
using the technique of phase dispersion 
minimization refined the photometric period to 1.1209 $\pm$ 0.0006 days.  We 
predicted and then observed an eclipse on
 14 May 2004 using 
the 1.2m telescope at the Fred L. Whipple Observatory, AZ, USA using
SLOAN filters $r$, $i$ and $z$.   
The long time base provided by this observation allowed us to refine 
the photometric period.
By fitting the lightcurves during (21) eclipse times
(we included only totally observed eclipses) to parabolas, we determined all 
the times of minimum light (center of eclipse)
with corresponding error.
For the 
eclipse observed on May 14th 2004, 
we only used the time of minimum light from the $r$ filter.  We also used
observations from IAC80 (see below).
Using the bootstrap method we refined the period to 1.12079 $\pm$ 0.00001
corresponding to a precision of 1 second.
The epoch of secondary minimum, T$_0$, was meanwhile determined to be
2453139.749509 (HJD) $\pm$ 0.000075.

TrES--Her0-07621   has a stellar neighbor at a distance of 8'', 
close enough that
the two objects are blended in our STARE observations (STARE has a pixel
size of about 11 arcsec). 
Observations in $R$ and $I$ Johnson filters using the IAC80 at 
Observatorio del Teide on 30 August 2004 provided a 
more realistic picture of the depth of one of the eclipses, while also
allowing us to confirm the photometric period.
We measured the PSF of both the binary and the neighbor using all five 
images outside of the eclipse.  
From these we derived the R fractional flux contribution from this 
companion star of 0.19 $\pm$ 0.04.
Because the companion star is also quite red (Table \ref{facts}), the flux should be
similar (to within the error) in both Johnson and Harris R filters, 
and so we can use this number to analyze the STARE time series.   
Measurement of
the contamination of the eclipse signal from TrES--Her0-07621 
by the companion star
is important, because it must be accounted for when fitting the
time-series data to estimate the stellar radii.
This neighbor also has a proper motion that is similar
in magnitude and direction to that of TrES--Her0-07621, indicating the 
possibility that TrES--Her0-07621 is at least 
a triple system, with
the eclipsing pair of stars accompanied by a third M dwarf at a distance
of hundreds of AU.

Adopting the photometric period as the orbital period and introducing its associated error,
we fit all 36 radial velocities (blue, red, H$\alpha$) with a Keplerian model 
using GaussFit 
(Jefferys, Fitzpatrick \& McArthur, 1988).  
The model is similar to that used in McArthur et al. (2004). We assume
an eccenctricity $e$ of 0.
The resulting radial velocity semi-amplitudes are
${\rm K}_1 = 100.54 \ \pm 0.31$ km s$^{-1}$ and 
${\rm K}_2 = 101.29 \ \pm 0.31$ km s$^{-1}$,
$({\rm M}_1 + {\rm M}_2) \sin^3 i = 0.9547 \ \pm 0.0062$ M$_\odot$,
and M$_1$/M$_2 = 1.0075 \ \pm 0.0044$.
A formal solution including eccentricity (e = 0.006 $\pm$ 0.002) provided a
better solution, reducing $\chi^2$ by 8\%, while reducing the number of degrees
of freedom by 3\%. 
However, we constrain $e$ = 0 for this analysis.

We developed a chi-square minimization algorithm
to estimate orbital parameters from the light curve, ignoring any variations
between eclipses (Figure  3).
The input parameters are period $P$, component masses M$_1$ and M$_2$, 
limb-darkening coefficients (0.7, Claret, 1998, Table 7) 
and the light from a third nearby
 star as a fraction of the total light of the system (0.19 $\pm$ 0.04).  
The code solves for both radii R$_1$ and R$_2$, effective temperature ratio 
T$_2$/T$_1$, 
center of minimum eclipse T$_0$ and 
inclination $i$.
Figure \ref{gr_fit_res} shows the resulting fit to the data.

The initial estimates for  R$_1$, R$_2$, T$_2$/T$_1$ and $i$ were derived from 
two-dimensional $\chi^2$ contour
plots (while keeping the other two parameters fixed).  
These contour plots presented high correlations between the two 
radii, constraining their sum while insensitive to their difference;
 and between radius (R$_1$ or R$_2$) and inclination; larger radius implies
smaller inclination. 
T$_2$/T$_1$ was uncorrelated to both radii and inclination, so its 
error is given by the corresponding value of this parameter at 
$\chi^2 + \sigma$ (Press et al, 1986) in the direction of its axis.   
However, because the other parameters are obviously
not independant, R$_1$ and R$_2$ for example, the error spanned the range of 
radii where the contour value is $\chi^2 + \sigma$, (the full range error ellipsoid).

Even with the component masses determined, in absence of a T$_{eff}$ 
measurement we require the component 
absolute magnitudes to place these stars on the Mass-Luminosity Relation (MLR).
From the TrES data,
calibrated using stars within 1$^{\circ}$ that have measured V magnitudes 
from SIMBAD,
we obtain a V-band apparent magnitude of 15.51 $\pm$ 0.08 for the combined
3-star system, giving
V-K = 4.63 $\pm$ 0.10 (Table \ref{facts}).  
Assuming a wavelength independent relative flux
we estimate $\Delta$V$_{AB-C}$ = 1.72 from difference J, H and K magnitudes
 (between the neighbor and the binary). 
We can also estimate $\Delta$V$_{A-B}$ (between the binary components) 
= 0.1 $\pm$ 0.05, 
based on the derived temperature and radii differences.
Taking all of the above into account, we estimate
 the component magnitudes of 
V$_A$ = 16.37 $\pm$ 0.1, V$_B$ = 16.56 $\pm$ 0.1
and V$_C$ = 17.43 $\pm$ 0.1, where $C$ is the stellar neighbor.

From Hawley et al, (2002) color-spectral type relations we estimate
an M3 spectral type for each component.
We obtain from the Hawley
$M_J$ - spectral type 
relationships component absolute magnitudes of 
M$_V$ = 11.18, 11.28 $\pm$ 0.3.
Accepting this estimate of the luminosities, 
the distance modulus is $\mu$ ($\sim$16.4-11.2) $\sim$ 5.2, 
corresponding to d$\sim$110 pc.  
For this nearby system we have assumed no absorption (A$_V$ = 0). 
We also use the radii and effective temperature (Table \ref{tbl-orbital}) to 
determine luminosities,
differentially with respect to the Sun (e.g. Benedict et al, 2003).
With  bolometric corrections as a function of temperature from
Flower (1996) we obtain an average d = 118 $\pm$ 13 pc for the
two components.

\section{Results and Comments}

Using all the derived parameters and errors, we refit the lightcurve using 
our code, and tested these results with the code 
{\it Nightfall}\footnote{http://www.lsw.uni-heidelberg.de/users/rwichman/Nightfall.html} 
(see below). 
Both codes give similar results, 
their difference being within the error bars.
Tables \ref{tbl-system} and \ref{tbl-orbital} summarize the results.

\begin{figure*}
\plotone{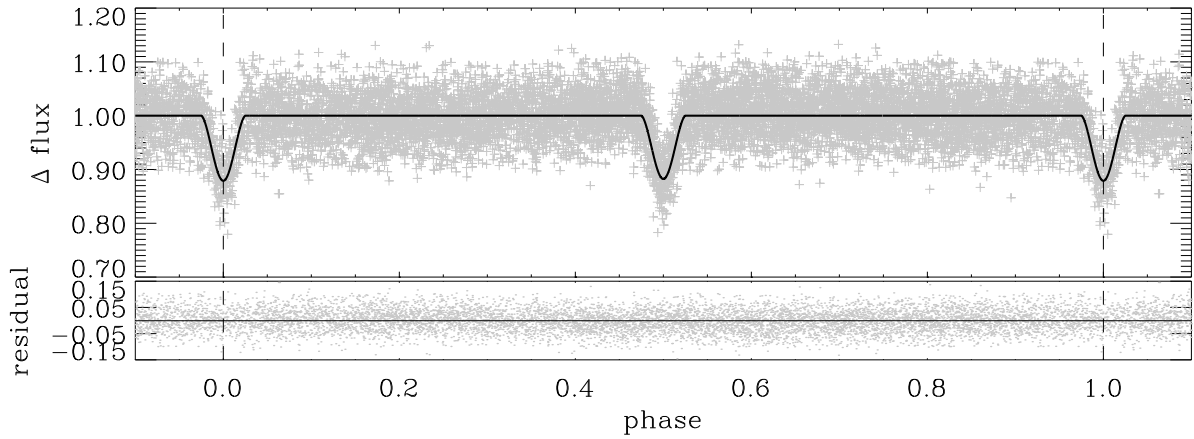}
\caption{Observed light curve and fitting results. 
The top panel shows the reconstructed lightcurve (continuous line)  using the 
parameters obtained by fitting the observed lightcurve without
the out-of-eclipse variations (small crosses, see section Results.).  We 
have indicated phase = 0 and 1 (center of secondary eclipse) 
by dashed lines.
The bottom panel shows the residuals (dots), with the continuous line 
representing the rms of residuals.}
\label{gr_fit_res}
\end{figure*}

Our code does not allow for stellar spots, so we subtracted a smooth function 
(by a Fourier technique) to remove the out-of-eclipse variations 
making a rectified lightcurve.
We also constrain $e$ = 0. 
The top panel of Figure \ref{gr_fit_res} shows the synthetic lightcurve
(continuous line) 
corresponding to the model fit (our code) of 
the folded light curve (small crosses). 
Phase = 0 corresponds to the secondary eclipse.
The bottom panel shows the residuals of the fit. 
The residuals show no variation as a function of phase indicating
an adequate model fit.

Because our code is unable to account for spot variability, we inspected the
residuals after subtracting the model fit from the unrectified lightcurve.  
These residuals also showed no evidence of eclipses. 
We also fit this unrectified lightcurve to find R$_1$, R$_2$, $i$, T$_0$ and 
T$_2$/T$_1$. 
The results varied slightly from those for the rectified light curve, 
but stayed within the 
error bars (Table \ref{tbl-orbital}).  

Our original (unrectified) photometric light curve contains non-uniform 
outside-eclipse variations.  Binary systems
 such as TrES--Her0-07621 are often magnetically active
 (e.g., Strassmeier et al. 1993).
  While tidal effects may be important, 
these non-uniform variations are most likely
 explained by star spots. 
We used {\it Nightfall}
to model our unrectified lightcurve, 
because this code allows for the presence of spots
 on each of the components.  Our derived parameters were used as
inputs and we attempted 
to solve for the longitude, latitude and radii of spot(s).
There was no unique solution; many combinations of these spot parameters
could compensate for the out-of-eclipse variations,
although they always presented a 180$^{\circ}$ longitude difference.
  This preferred longitude 
 difference has also been observed in 
other active binary systems (see e.g., Henry et al. 1995).
The presence of spots can have a significant effect on the 
accuracy of the derived parameters, such as inclination, temperature and radii
(Torres \& Ribas (2002) discuss this for the case of YY Gem). 
Additional observations, photometry 
in particular, will be necessary to increase the precision of the 
radii estimates as well as to learn more about the magnetic behaviour of the
stars. 
This could then 
provide a link towards a better understanding of the physical processes 
of these low-mass objects.

\acknowledgments
We thank Hect\'or Vazquez Ramio (observations on IAC80),
and operating staff for STARE.  The IAC80 and STARE 
are operated by the Instituto 
de Astrof\'isica de Canarias in
the Spanish Observatorio del Teide. 
We thank Mark Everett (observations on 48-inch telescope at 
Fred L. Whipple Observatory on Mount Hopkins, Arizona USA,
operated by Harvard-Smithsonian Center for Astrophysics.)  
We also thank Rainer Wichmann for the use of the program 
for the light-curve synthesis, {\it Nightfall}.
Support for this work was provided by NASA through grants 
GO-09408 and GO-09407 the Space Telescope Science Institute, 
which is operated by the Association of Universities of Research in Astronomy, 
Inc., under NASA contract NAS5-26555.  
The Hobby-Eberly Telescope (HET) is a joint project of 
the University of Texas at Austin, the Pennsylvania State University, 
Stanford University, Ludwig-Maximilians-Universitat Muenchen, 
and Georg-August-Universitat, Goettingen.  
We thank the HET resident astronomers and telescope operators.
We thank the referee for their constructive comments.

\clearpage

\begin{table*}
\begin{center}
\caption{Catalog Information \label{facts}}
\begin{tabular}{lcc}
 & Binary & Neighbor\\
\tableline
J\tablenotemark{a}  & 11.773 & 13.487\\
H &  11.137 & 12.863\\
K & 10.880 & 12.615\\
J-K & 0.893 & 0.872\\
$\mu_{\alpha}$\tablenotemark{b} & -2 $\pm$ 8 & -26 $\pm$ 14\\
$\mu_{\delta}$ & +30 $\pm$ 3 & +28 $\pm$ 7\\
\tableline
\tablenotetext{a}{2MASS catalog}
\tablenotetext{b}{USNO-B catalog, units mas yr$^{-1}$}
\end{tabular}
\end{center}
\end{table*}

\clearpage

\begin{table*}
\begin{center}
\caption{System Parameters \label{tbl-system}}
\begin{tabular}{lcrcl}
\tableline

P(days)&\hspace{0.5cm} & 1.12079 & $\pm$& 0.00001\\
P(years)&&0.00306861 &$\pm$& 0.00000036\\
T$_0$(JHD)&& 2453139.749509 & $\pm$ & 0.00075\\   
M$_{T}$(M$_{\odot}$)
 & & 0.983& $\pm$& 0.007\\
a(AU) && 0.01047 & $\pm$& 0.00002\\
a(R$_{\odot}$)  && 2.251 & $\pm$ & 0.005\\
$i$($^{\circ}$) && 83.12& $\pm$& 0.30\\
$\gamma$(km s$^{-1}$) && -26.5 & $\pm$ & 0.3\\
T$_B$/T$_A$  && 0.97 &$\pm$ &0.02 \\

\tableline
\end{tabular}
\end{center}
\end{table*}

\clearpage

\begin{table*}
\begin{center}
\caption{Component Parameters \label{tbl-orbital}}
\begin{tabular}{lcrclcrclrcrclcrcl}
 & &   & A & && & B \\
\tableline

M(M$_{\odot}$) &\hspace{1cm}&
0.493 & $\pm$ & 0.003  & \hspace{3mm}& 0.489 & $\pm$ & 0.003  &\\
M$_V$ & & 11.18 & $\pm$  &0.30 && 11.28 & $\pm$ &0.30\\
K(km s$^{-1}$) & & 
100.54& $\pm$  &0.31 && 101.29 &$\pm$ & 0.31\\ 
R(R$_{\odot}$) && 
0.453 &$\pm$ & 0.060  && 0.452 &$\pm$ & 0.050 \\
T$_{eff}$\tablenotemark{a} (K) & & 3500 & &  && 3395 & &  \\
\tableline
\tablenotetext{a}{The component A temperature is based on that expected of an M3V star (Cox, 2000).}
\end{tabular}
\end{center}
\end{table*}

\clearpage

\clearpage

\clearpage

\end{document}